\documentclass[runningheads]{llncs}
\usepackage[T1]{fontenc}
\usepackage{graphicx}
\usepackage{booktabs}
\usepackage[misc]{ifsym}
\newcommand{\corr}{(\Letter)}
\usepackage{mwe}
\usepackage{subcaption}
\usepackage{tabularx}
\usepackage{enumitem}
\usepackage{amsmath}
\usepackage{caption}

\begin{document}

\title{Personalized Contest Recommendation in Fantasy Sports}

\author{Madiraju Srilakshmi \corr  \and Kartavya Kothari   \and Kamlesh Marathe \thanks{work done while at Dream11} \and Vedavyas Chigurupati \and Hitesh Kapoor 
}
\authorrunning{M.Srilakshmi et al.}

\institute{Dream11, India \email{\{madiraju.srilakshmi,kartavya.kothari, vedavyas.chigurupati, hitesh.kapoor\}@dream11.com, kamleshmaratheiitb@gmail.com}
}
\toctitle{Personalized Contest Recommendation in Fantasy Sports}
\tocauthor{Madiraju Srilakshmi, Kartavya Kothari, Kamlesh Marathe, Vedavyas Chigurupati, Hitesh Kapoor }

\maketitle              

\begin{abstract}
In daily fantasy sports, players enter into ``contests'' where they compete against each other by building teams of athletes that score fantasy points based on what actually occurs in a real-life sports match. For any given sports match, there are a multitude of contests available to players, with substantial variation across 3 main dimensions: entry fee, number of spots, and the prize pool distribution. As player preferences are also quite heterogeneous, contest personalization is an important tool to match players with contests. This paper presents a scalable contest recommendation system, powered by a Wide and Deep Interaction Ranker (WiDIR) at its core. We productionized this system at our company, one of the large fantasy sports platforms with millions of daily contests and millions of players, where online experiments show a marked improvement over other candidate models in terms of recall and other critical business metrics.

\end{abstract}

\section{Introduction}
Over the past decade, fantasy sports have grown remarkably. The global market is estimated to exceed \$37 billion in 2025 and is projected to continue to grow by 14\% annually through 2030 \cite{mordor_fantasy_sports}. The majority of this growth is driven by daily fantasy sports (DFS), where players create teams of athletes and compete against each other on a match-by-match basis. To enter into these competitions--typically referred to as ``contests''--players pay an entry fee which contributes to a shared prize pool. Once a contest concludes, this prize pool is then divided across the top-performing entries according to a predetermined prize distribution.

In this paper, we focus on the contest recommendation problem at our company, one of the large fantasy sports platforms globally with millions of players. For a real-life sports match, our platform hosts a wide variety of different contests for players to join, ranging from small head-to-head clashes to extremely vast competitions that involve millions of teams. Beyond just size (number of spots), contests also principally differ across two other key dimensions: entry fees and prize distributions. Player preferences are also quite diverse, meaning that personalized recommendations can be critical in directing players to contests that best align with their tastes, and can generate substantial lifts in player conversions and entry amounts for the platform \cite{chugh2023ready}.

To solve this problem, we built what we call a Wide and Deep Interaction Ranker (WiDIR), which builds upon the architecture developed to rank apps on the Google Play store \cite{cheng2016wide}. WiDIR similarly integrates a wide linear branch, which excels at ``generalization'', and a deep neural network branch, which specializes in ``memorization''. 

In the context of fantasy sports, generalization in contest recommendation refers to the ability to recommend relevant contests to new players or recommend newer or less popular contests that may still be a good fit to the existing players. Memorization, however, focuses on recommending popular contests to players who have previously shown a preference for them. Our recommendation model is designed to achieve both of these capabilities.

We further enhance the deep network by explicitly modelling contest-player interactions and by dividing it into three specialized sub-branches: one dedicated to player features, another to contest features, and a third focused specifically on interaction features. Our offline experiments show that WiDIR achieves better precision and recall compared to other candidate models on out of sample test data. Moreover, online experiments showed that significantly more lift on player engagement and other key business metrics against other candidates.

The primary contributions of our paper are listed below;
\begin{itemize}
    \item We introduce WiDIR (Wide and Deep Interaction Ranker), a personalized recommendation system specifically tailored for daily fantasy sports (DFS) contests. 
    
    
    \item Through extensive experimentation on a large-scale DFS platform involving millions of players, WiDIR achieves a significant uplift in key player engagement and business metrics such as contest joins and gross gaming revenue (GGR).
    \item We present a fully operational and scalable inferencing pipeline designed to deliver personalized contest recommendations within milliseconds, effectively handling millions of players and contests daily.
\end{itemize}

The rest of the paper is organized as follows; Section \ref{sec:context} provides additional context on fantasy sports as well as a brief overview of our platform. Section \ref{sec:datamethod} presents our data and methodology. Section \ref{sec:system} describes our overall system. Section \ref{sec:experiments} discusses the experimental results and analysis. Finally, Section \ref{sec:conclusion} concludes the paper.

\section{Context}
\label{sec:context}
\begin{figure}[t]
\label{fig:D11app}
\centering
\begin{subfigure}{.5\textwidth}
  \centering
  \includegraphics[width=.5\linewidth]{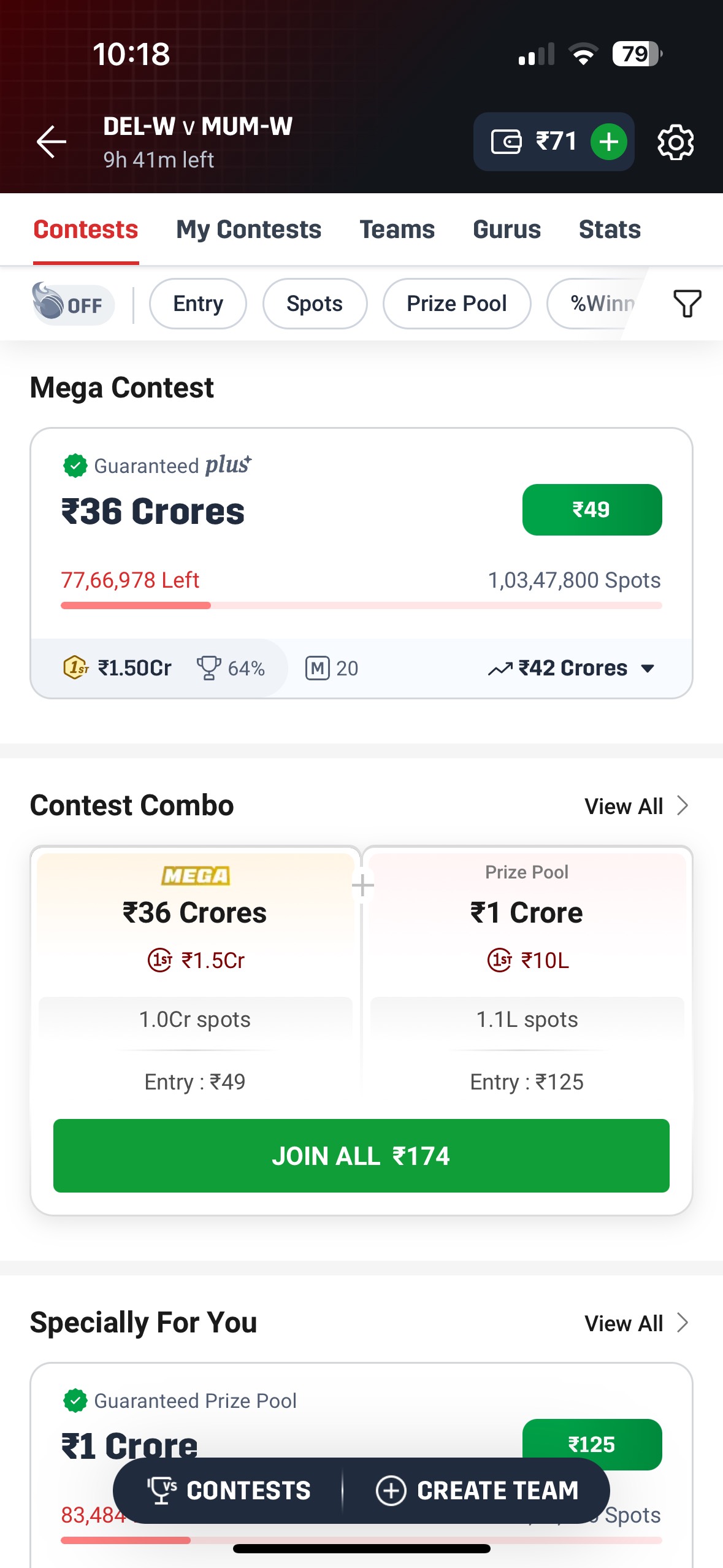}
  \caption{Contest categories listed}
  \label{fig:mega_bundle}
\end{subfigure}%
\begin{subfigure}{.5\textwidth}
  \centering
  \includegraphics[width=.5\linewidth]{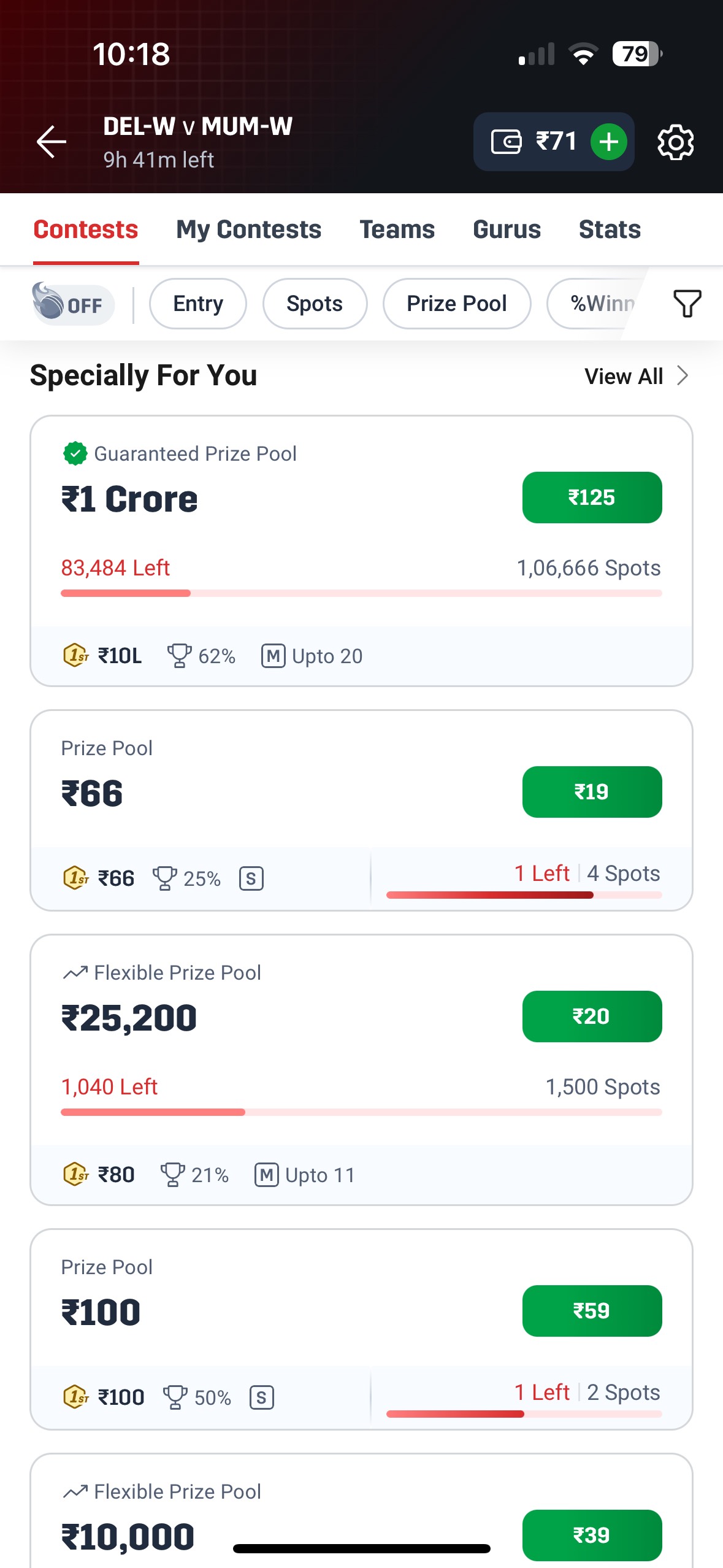}
  \caption{\textit{Specially For You} Section}
  \label{fig:sfy_ss}
\end{subfigure}
\caption{Some of the contests on our platform for Delhi (DEL-W) vs Mumbai (MUM-W) TATA WPL match}
\label{fig:test}
\end{figure}

Fantasy sports involve creating virtual teams composed of real-life athletes, where the performance of these athletes in actual sporting events determines the success of the fantasy team. Players draft athletes, manage rosters, and compete based on the accumulated statistical metrics of their athletes, such as runs scored in cricket, touchdowns in football, or goals in soccer. These metrics have an equivalent ``fantasy score'', for example in cricket: 
\begin{itemize}
    \item Run Scored: 1 point per run
    \item Wicket Taken: 25 points per wicket
    \item Catch: 8 points per catch
    \item Stumping/Run-out: 12 points per dismissal
    \item Boundary Bonus: 1 point for each boundary (4 runs), 2 points for each six (6 runs)
\end{itemize}

Traditional fantasy sports competitions last an entire season, requiring long-term strategic planning, regular engagement, and active management of team rosters. However, in recent years, daily fantasy sports have become increasingly popular due to their shorter contest durations, typically spanning a single day. Unlike traditional fantasy sports, where teams remain relatively stable throughout a season, DFS participants create new teams for each contest. DFS’s appeal lies primarily in its immediate outcomes and reduced time commitment, attracting a broader range of participants. On our platform, after entering into a contest, players will construct a team from the athletes who are competing in the corresponding real-life sports match. 

We host three types of contests: Public, Special, and Mega (or Grand). ``Public contests'' are non-guaranteed, with spots ranging from 2 to over 1,500. These contests are regenerated each time an instance fills up. ``Special contests'' offer guaranteed prize pools, often featuring highly attractive top-prize templates. These contests provide significant business benefits and are among the most appealing on the platform. ``The Mega (or Grand) contest'' is a special contest with the largest prize pool, standing out as the most prestigious and high-value contest available.

Our proposed WiDIR excels in Daily Fantasy Sports (DFS) by effectively leveraging dynamic player-contest interactions and rapidly adapting to evolving player preferences. Unlike traditional season-long fantasy sports, DFS's highly dynamic contest environment and immediate feedback loop allow WiDIR to continuously adapt recommendations, utilising interaction features to enhance player engagement and business metrics.

Figure \ref{fig:mega_bundle} depicts the player interface after selecting a specific cricket match (Delhi (DEL-W) vs Mumbai (MUM-W) TATA WPL match) on our mobile app. At the top is the ``Mega Contest'', the largest and most lucrative contest, offering significant prize pools that attract up to millions of players. Directly below are contest bundles, known as ``Combo Contests''. allowing players to conveniently join multiple contests simultaneously at discounted entry fees, enhancing the overall playing experience and value proposition.

Further down, players encounter the ``Specially For You'' section, illustrated in greater detail in Figure \ref{fig:sfy_ss}. This section leverages advanced personalization algorithms, recommending contests tailored to individual player preferences and historical playing behaviour, thereby increasing engagement and player satisfaction. As much of the rest of the first page is more business-optimized, this section is explicitly optimizing for player engagement, ensuring a more balanced experience. Players benefit from organic engagement driven by personal preferences while maintaining visibility for key business offerings.

\subsection{Problem Definition}
Our contest recommendation model aims to predict which contests a user is most likely to join in future rounds, leveraging their historical join behavior and the features of available contests. 

Our primary objective is to generate personalized contest recommendations that align closely with individual player preferences. By improving the relevance of contest recommendations, we aim to enhance user engagement, increase contest participation, and ultimately support long-term player retention and platform revenue growth.

\section{Data and Methodology}
\label{sec:datamethod}
\subsection{Primary Data and Feature Engineering}
\label{subsec:data_and_FE}

Our primary data comprises two main components: the player contest join history and contest characteristics. The join history records contain \texttt{player\_id}, \texttt{contest\_id}, \texttt{match\_id}, and \texttt{joining\_time}, while the contest characteristics include \texttt{entry\_fee}, \texttt{prize\_money}, \texttt{contest\_size}, \texttt{prize\_distribution}, \texttt{guaranteed}, \texttt{contest\_type}, and \texttt{multi\_entry}. 

Among these, \texttt{contest\_size}, \texttt{contest\_type}, \texttt{prize\_distribution}, \texttt{guaranteed}, and \texttt{multi\_entry} are categorical variables, whereas \texttt{entry\_fee} and \texttt{prize\_money} are numerical. Below is a brief description of each feature:
\begin{itemize}
    \item \texttt{player\_id}: Unique identifier for each player.
    \item \texttt{contest\_id}: Unique identifier for each contest. Multiple instances of the same contest can exist; once one contest is filled, a new contest with identical features is dynamically created and assigned a different \texttt{contest\_id}.
    \item \texttt{match\_id}: Unique identifier for the sports event associated with the contest.
    \item \texttt{joining\_time}: Timestamp indicating when a player joined a contest.
    \item \texttt{entry\_fee}: Amount required to participate in a contest.
    \item \texttt{prize\_money}: Total prize pool for a contest.
    \item \texttt{contest\_size}: Maximum number of participants allowed in a contest.
    \item \texttt{contest\_type}: Type of contest (e.g., public, special, mega). 
    \item \texttt{prize\_distribution}: Template identifier representing the prize distribution among winners.
    \item \texttt{guaranteed}: Indicates whether the contest is guaranteed (i.e., the prize money remains fixed even if the contest is not fully filled).
    \item \texttt{multi\_entry}: Indicates whether a contest allows multiple team entries from the same player.
\end{itemize}

From our primary data, we engineered three distinct sets of features: player features, contest features, and interaction features.
\begin{itemize}
    \item \textbf{Player Features}: These features characterize a player's behaviour based on their historical contest participation. For each player, we compute statistics over the contests joined in the last \( k \) days. The goal is to capture current player behavior (\(k=3\)) alongside historical preferences (\(k \in (7,30)\)).
    
    The features include:
    \begin{itemize}[label={--}]
        \item \textit{Contest Join Behavior}: Counts of distinct contest types, contest sizes, and entry fees encountered.
        \item \textit{Spending Patterns}: Average and maximum entry fees paid.
        \item \textit{Winning Proportion}: Average and maximum prize money won.
        \item \textit{Diversity Level}: Count of distinct types of contests played.
    \end{itemize}
    \item \textbf{Contest Features}: These features describe the intrinsic attributes of a contest, as outlined above.
    \item \textbf{Interaction Features}: These capture the current affinity between a player and a contest by aggregating player behavior over recent time windows. Specifically, we compute features over the contests the player participated in over the previous day and the last \( k \) days (with \( k=5\)). 
    
    Examples include the count of contests with the same type as the target contest, count of contests with the same entry fee as the target contest, and number of contests with the same prize pool as the target contest.
\end{itemize}

This final dataset is then split into training (\(\sim\) 12 months), validation (\(\sim\) 2 months) and test sets (\(\sim\) 6 months). Table \ref{table:d11_data_stats} summarizes key data statistics, including the duration of the training and test periods, contest join counts, and the number of unique players and contests.

\begin{table}[t]
\centering
\begin{tabularx}{0.75\linewidth}{|X|X|X|}
\toprule
\textbf{Property} & \textbf{Train} & \textbf{Test} \\
\midrule
Contest joins & \(\sim\) 1 Billion & \(\sim\) 0.5 Billion \\
Total players & \(\sim\) 100 thousand & \(\sim\) 100 thousand \\
Unique contests & \(\sim\) 1.5 thousand & \(\sim\) 1.1 thousand \\
Total contests & \(\sim\) 1.5 Million & \(\sim\) 0.9 Million \\
\bottomrule
\end{tabularx}
\caption{Train and Test Data Statistics}
\label{table:d11_data_stats}
\end{table}

\subsection{Methodology}
\label{subsec:method}
Our contest recommendation system aims to predict the ranking of contests that a player is most likely to join in future matches, given their contest transaction history. Let $i$ index the set of players $\mathcal{P}$, $t$ index the set of matches $\mathcal{M}$, and $c$ index the set of contests $\mathcal{C}$. We construct training samples at the player-match level as the active set of contests varies from match to match: for each player $i \in \mathcal{P}$ and match $t \in \mathcal{M}$, we construct an ordered list of contests:
\[
\mathcal{O}_{it} = \Bigl(c_{it}^{(1)}, c_{it}^{(2)}, \dots, c_{it}^{(L_{it})}\Bigr)
\]
where \( c_{it}^{(1)} \) denotes the most frequently joined contest by player $i$ in match $t$ to \( c_{it}^{(L_{it})} \) which denotes the least. This list is then fixed to 100 entries, either trimmed down or padded by randomly appending contests that weren't joined to the end of $\mathcal{O}_{it}$. We set this value based on our experiments with \((50,100,200)\). From $\mathcal{O}_{it}$, we can form the set of contest pairs:
\[
\Theta_{it} = \Bigl\{ (c, c') \in \mathcal{O}_{it} \times \mathcal{O}_{it} \mid c \succ c' \Bigr\},
\]
where \( c \succ c' \) indicates that contest \( c \) was joined more frequently than contest \( c' \) by \( u_i \). These pairs are constructed to train our ranking model, WiDIR using a pairwise hinge loss \cite{joachims2002optimizing}. A ranking loss was chosen since we aim to rank the contests based on their relative relevance rather than accurately predicting the affinity score of each contest. 

WiDIR produces an affinity score \( \hat{s}_{ic} \) for each player $i$, each match $t$, and each contest $c$. This model is then trained using pairwise hinge loss as mentioned above.

\begin{align} \label{eq:loss_pairwise}
\mathcal{l}_{it}(c, c') = \max\{0, \; 1 - \hat{s}_i(c) + \hat{s}_i(c')\}.
\end{align}
which penalizes situations where a less preferred contest \( c' \) is scored higher than a more preferred contest \( c \). The overall loss is given by
\begin{align}\label{eq:total_loss}
\mathcal{L} = \sum_{i \in \mathcal{P}} \sum_{t \in \mathcal{M}} \sum_{(c, c') \in \mathcal{O}_{it}} \mathcal{l}_{it}(c, c').
\end{align}

\begin{figure}[t]
    \centering
    \includegraphics[width=0.8\linewidth]{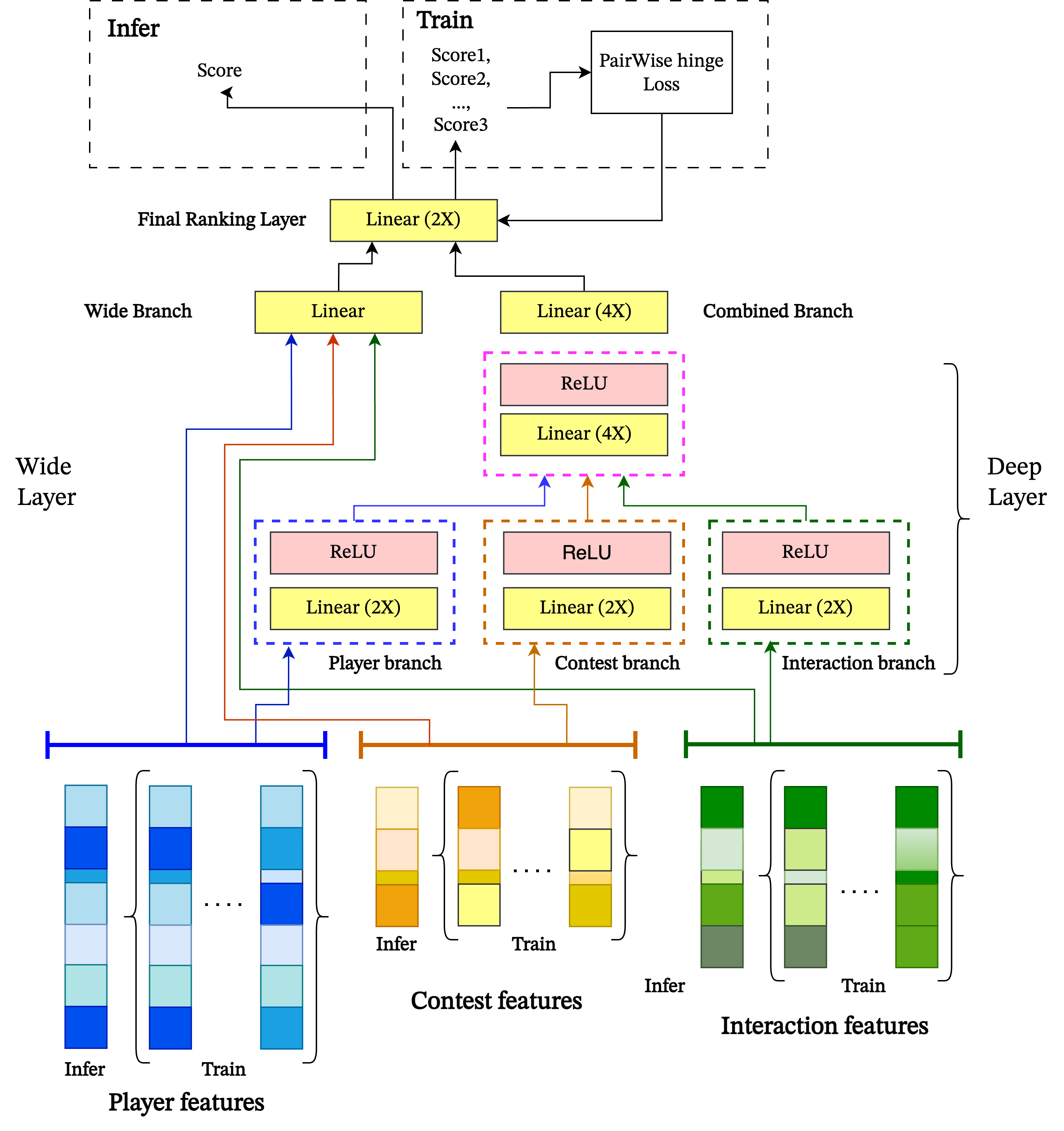}
    \caption{WiDIR architecture showing both training and inference flow}
    \label{fig:model_arch}
\end{figure}

WiDIR is built on a wide and deep neural architecture to leverage both memorization and generalization capabilities (see Figure \ref{fig:model_arch}). Unlike \cite{cheng2016wide}, our model employs three separate deep branches to model the embeddings for players, contests, and their interactions. This approach allows each component to be represented more effectively. The embeddings are then combined in a final branch to model the correlations between them. The training process is described below:
\begin{itemize}
    \item Embed the player, contest, and interaction features into dense representations.
    \item Concatenate these embeddings and pass them through multiple fully connected layers (the deep component).
    \item In parallel, combine the raw features into a wide layer.
    \item Merge the wide and deep components, then feed the result into a final linear layer to predict contest scores.
    \item Compute the pairwise hinge loss from Equation \eqref{eq:loss_pairwise} over all contest pairs and back-propagate the error.
\end{itemize}
\begin{table}[hbt!]

    \centering 
    \begin{tabular}{|l|c|c|}
        \hline
        \textbf{Component} & \textbf{Layer Flow} & \textbf{Parameter Count} \\
        \hline
        Player Branch (PB) & \( P \rightarrow 64 \rightarrow 64 \) & \( 64 * d_p + 4224 \) \\
        \hline
        Contest Branch (CB) & \( C \rightarrow 64 \rightarrow 64 \) & \( 64 * d_c + 4224 \) \\
        \hline
        Interaction Branch (IB) & \( I \rightarrow 16 \rightarrow 16 \) & \( 16 *d_i + 560 \) \\
        \hline
        Wide Branch (WB) & \( P+C+I \rightarrow 1 \) & \( d_p + d_c + d_i + 1 \) \\
        \hline
        Deep Branch (DB) & \( (64+64+16) \rightarrow 128 \) (4X) & 68096 \\
        \hline
        Combined Layer & \( 128 \rightarrow 64 \rightarrow 64 \rightarrow 32 \rightarrow 8 \rightarrow 4 \) & 14796 \\
        \hline
        Final Ranking & \( (4+1) \rightarrow 4 \rightarrow 1 \) & 29 \\
        \hline
        \textbf{Total} & & \(\mathbf{65 * d_p + 65 * d_c + 17 * d_i + 91930}\) \\
        \hline
    \end{tabular}
    \caption{Model Architecture and Parameter Counts (Symbolic)}
    \label{tab:model_architecture}
\end{table}
Table \ref{tab:model_architecture} summarizes the layer flow and parameter counts in each branch. Here, \( d_p \), \( d_c \), and \( d_i \) denote the dimensionalities of the player, contest, and interaction features, respectively (our raw feature set dimensions were 107, 11, and 9 in our experiments).

During inference, the model scores all contests available in a match based on their features. The top-\( h \) contests, as determined by their predicted scores, are then recommended to the player. Since the model relies on high-level contest features (e.g., \texttt{contest\_type}, \texttt{entry\_fee}) rather than contest-specific identifiers, it can generalize effectively to new contests.

\section{System Architecture and Data Flow}
\label{sec:system}

\begin{figure}[t]
    \centering
    \includegraphics[width=\linewidth]{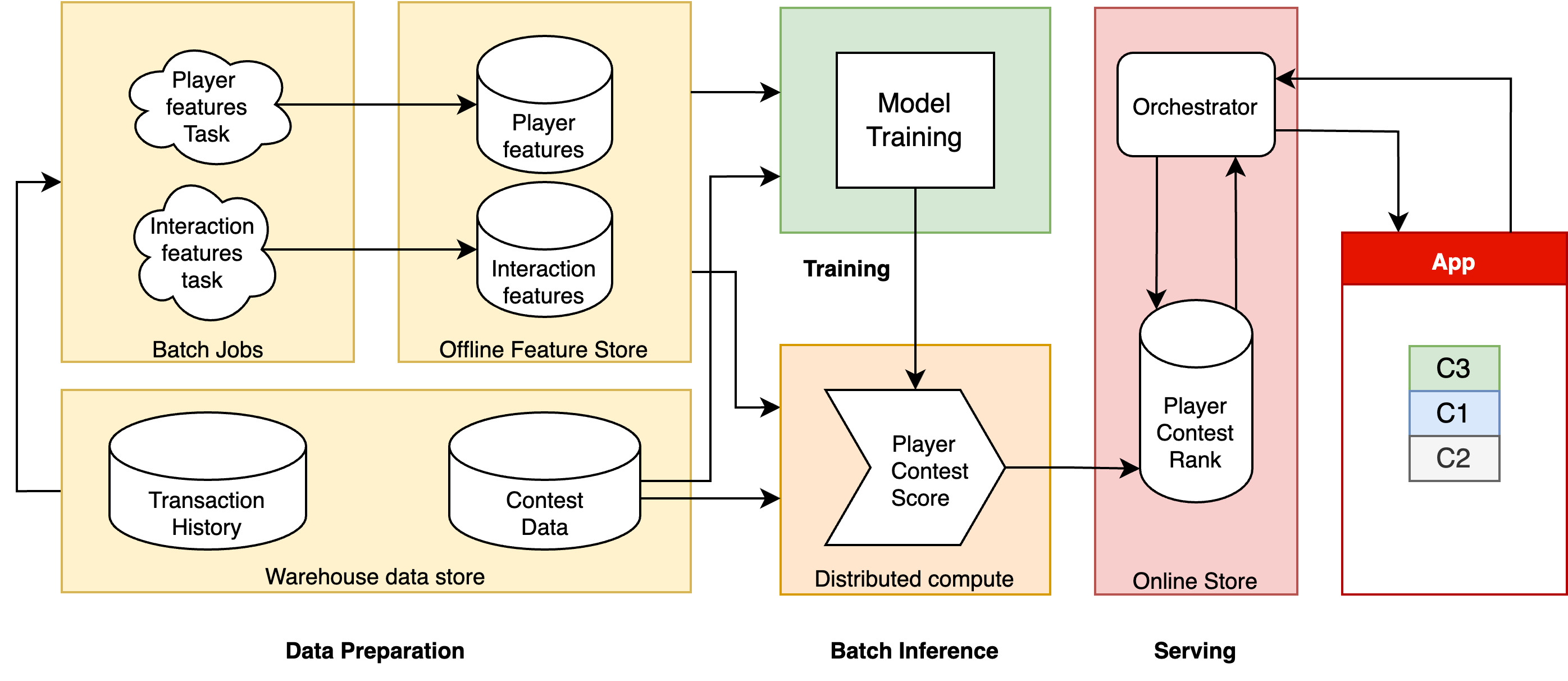}
    \caption{Data flow and Inference architecture}
    \label{fig:system}
\end{figure}

This section describes the architecture of the personalized contest recommendation system implemented in our fantasy gaming platform. 
The system integrates the following key phases as shown in Figure \ref{fig:system}. The following sections provide details on each phase.
\begin{enumerate}
    \item Data Preparation
    \item Training
    \item Inference
    \item Serving
\end{enumerate}

\subsection{Data preparation} 

We use player transaction history and contest features to generate our recommendations. This data is stored in a data warehouse which is maintained and gets updated with any real-time changes via data engineering stream/ETL pipelines.

We fetch the data from the warehouse and compute player features,  and interaction features using distributed pyspark jobs viz. \textit{Player features task} and \textit{Interaction features task}. Section \ref{subsec:data_and_FE} describes our feature engineering. 

This data is then processed into model-expected formats and stored in the \textit{Offline Feature Store}, serving as a repository for the processed feature data. Features are created and updated once a day since our feature aggregation granularity is day-level.

\subsection{Training} 

For model training, \textit{player features} and \textit{interaction features} are fetched from the offline feature store, and  \textit{contest features} are fetched from the data warehouse. Our baseline system, Light GBM ranker, is trained using synapseML \footnote{\url{https://microsoft.github.io/SynapseML/}} to efficiently process large datasets stored in PySpark dataframes. For WiDIR, we use tensorflow \cite{tensorflow2015-whitepaper} with petastorm data loaders \footnote{\url{https://github.com/uber/petastorm}}. The pair wise loss implementation is sourced from tensorflow-ranking.

The entire training process is managed through run identifiers enabling us to reproduce any component specific to a given run. We track those run-ids, model artefacts, hyperparameters, data statistics and evaluation metrics using MLflow \cite{zaharia2018accelerating} \footnote{\url{https://mlflow.org/docs/latest/}}.

\subsection{Batch inference}
During inference also, we fetch \textit{player features} and \textit{interaction features} from the offline feature store, and  \textit{contest features} from the data warehouse and input them to our \textit{trained ranking model}. The model outputs a player-contest affinity score for all available contests and we rank them in decreasing order of their scores. 

To optimize the balance between providing recommendations and minimizing the cost of running inference on players who may not engage with the app, inference is performed only for players who have been active in the past month.

\subsection{Serving}

To ensure that player rankings remain dynamic with contests currently live, we maintain a relative ranking of all available contests in our Sylla DB-based Online Feature Store. The orchestration layer performs the following operations: 

\begin{itemize}
    \item Whenever a player is active on the app and selects a gaming match, it gathers all live contests and requests the online feature store for the relative ranking of these contests. 
    \item The ranking is then served back to the app, all within 10 milliseconds. Each subsequent refresh or update, including specific triggers such as a contest being filled and needing replacement, involves a fresh call from the orchestrator.
\end{itemize}

\section{Experiments}
\label{sec:experiments}

To select the optimal modelling strategy, we employ a two-stage experimentation process that includes both offline and online A/B testing. This approach ensures thorough testing before production deployment. In this section, we outline our evaluation metrics, the baselines used for comparison, the experimental setup, and the results from both online and offline settings. 

\subsection{Baseline systems}
We compare WiDIR with the following baseline systems.
\begin{itemize}
    \item ML-based model: tree-based Light GBM ranker (LGB) \cite{ke2017lightgbm} with same input features as WiDIR. 
    \item Popularity-based ranker: This approach ranks contests by the prize amount.
\end{itemize} 

\subsection{Evaluation Metrics}
Our offline experiments are evaluated using precision@$h$ and recall@$h$, where $h \in (1,3,5,10)$. 
\begin{itemize}
    \item []\[
\text{Precision@h} = \frac{|\text{recommended contests@h} \cap \text{actual contests joined}|}{h}
\]
\item []\[
\text{Recall@h} = \frac{|\text{recommended contests@h} \cap \text{actual contests joined}|}{\text{|actual contests joined|}}
\]
\end{itemize}

We evaluate online simulations / AB tests using the following gameplay business metrics. Each metric is aggregated and then we compute the delta between the control group $CG$ and each target group $TG_i$ for the entire test period.
\[
    \Delta = \frac{M_{TG_{post}}-M_{CG_{post}}}{M_{CG_{post}}}-\frac{M_{TG_{pre}}-M_{CG_{pre}}}{M_{CG_{pre}}}
\]
Where $\Delta$ is delta and $M$ is the aggregated metric, one of the below. $CG_{pre}$ and $TG_{pre}$ are the pre-treatment start aggregates, and $CG_{post}$ and $TG_{post}$ are the post-treatment start aggregates.
\begin{itemize}
    \item CJ - denotes the total \textit{contest joins}. 
    \item CEA - denotes the cumulative \textit{entry amounts}.
    \item GGR - denotes the \textit{gross gaming revenue}.
\end{itemize}

\subsection{Experimental Setup}
The process starts with trained models viz. tree-based light gbm, WiDIR. Table \ref{tab:model_config} reports the details of hyperparameters for both systems. The parameter tuning was conducted using a hold-out validation set.
\begin{table}[t]
    \centering
    
    \begin{tabular}{|l|c|c|}
        \hline
        \textbf{Hyperparameter} & \textbf{WiDIR} & \textbf{LGB ranker} \\
        \hline
        Embedding Dimensions & 64 & - \\
        Epochs & 100 & 30 \\
        Learning Rate (LR) & 0.001 & 0.1 \\
        Batch Size & $2^{12}$ (4096) & 64 \\
        Validation Batch Size & $2^{14}$ (16384) & - \\
        Loss Function & pw\_hinge & lambdarank (application) \\
        Ranking Sort Order & - & player\_id, match\_id, join\_count \\
        Boosting & - & dart \\
        Num Leaves & - & 64 \\
        Early Stopping rounds & 15 & 10 \\
        Eval at & - & {1, 3, 5, 10} \\
        \hline
    \end{tabular}
    
    \caption{Hyperparameter Configurations for WiDIR and LGB Models}
    \label{tab:model_config}
\end{table}

\begin{table}[t]
    \centering
    
    \begin{tabular}{|c|c|c|c|}
        \hline
        \textbf{Group} & \textbf{Treatment} & \textbf{player Count} & \textbf{Duration} \\
        \hline
        CG    & None    & 1M & 6 Weeks \\
        $TG_1$ & Popular & 1M & 6 Weeks \\
        $TG_2$ & LGB     & 1M & 6 Weeks \\
        $TG_3$ & WiDIR   & 1M & 6 Weeks \\
        \hline
    \end{tabular}
    
    \caption{Experiment Details: Here $TG_i$ refers to the different cohorts on which we exposed the mentioned treatment. The cohorts were sampled from monthly active players, stratified randomly}
    \label{tab:experiment_details}
\end{table}
\subsubsection{Stage 1: Offline Experiments} 

We evaluated modelling strategies through offline experiments on a year's worth of player transactional data. The recommendation models were evaluated using precision and recall. 

\subsubsection{Stage 2: Online A/B Testing} 
Promising models based on offline results undergo A/B testing and deployed to a subset of players while a control group gets no treatment. We record player interactions and business impact before full deployment and record the business metrics after the experiment. Refer to Table \ref{tab:experiment_details} for details.

\subsection{Results and Analysis}
\begin{figure}[t]

    \centering
    \includegraphics[width=1.0\linewidth]{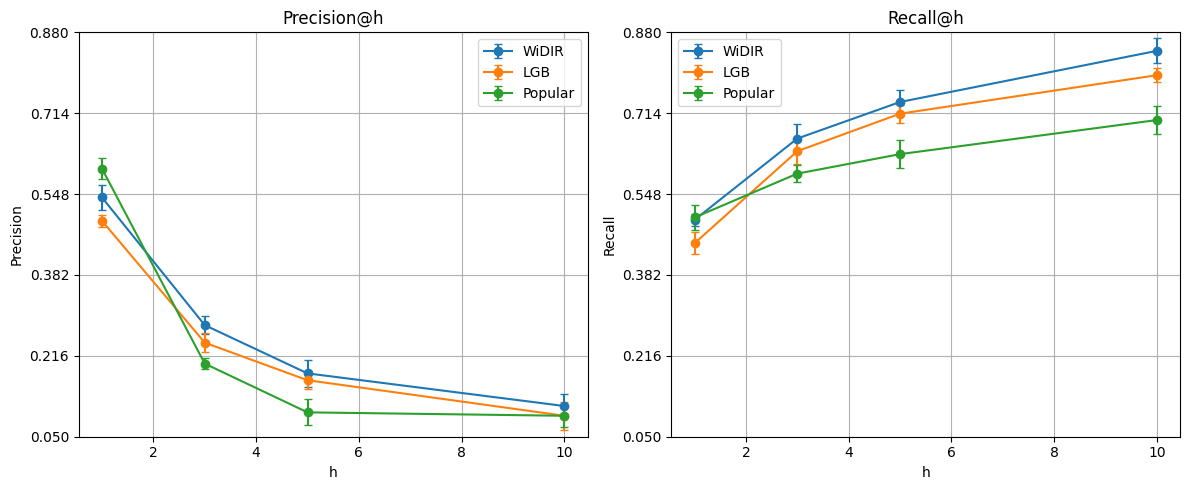}

    \caption{Comparing Precision@h and Recall@h of WiDIR against popular and LGB ranker}
    \label{fig:prec_recall_k}
\end{figure}

\begin{figure}[hbt!]

    \centering
    \includegraphics[width=0.7\linewidth]{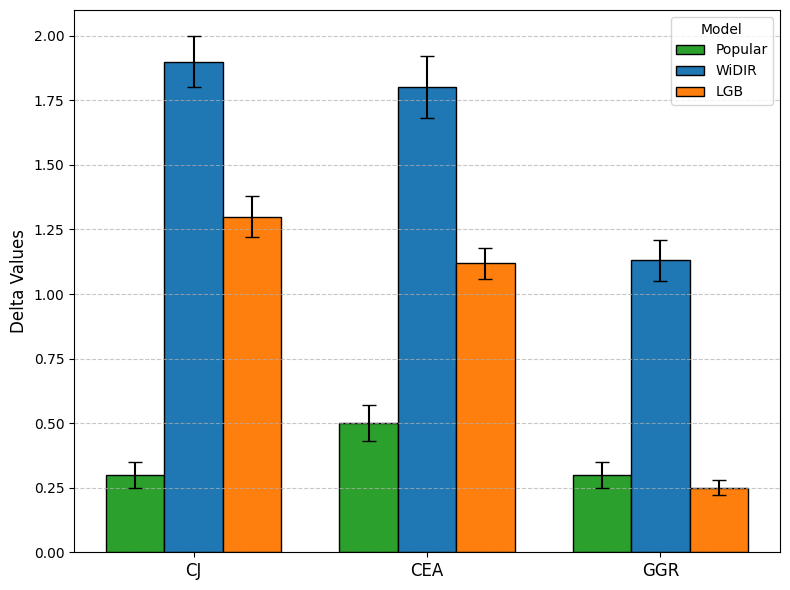}

    \caption{Comparing our online metrics (CJ delta, CEA delta and GGR delta) }
    \label{fig:online_radar_chart}
\end{figure}

The online and offline results of our wide and deep recommendation model and the baselines are shown in Figure \ref{fig:prec_recall_k} and Figure \ref{fig:online_radar_chart}. From the results, it is evident that WiDIR performs better than popular and LGB ranker in all but one case. The popular performs better at position 1 as the most popular contest is one that most players join anyway as it is attractive and strategically priced.

Offline experiments indicate that, as number of recommended contests ($h$) increases, precision decreases while recall increases for all models. This is because most players join only one or two contests in a given match, fewer than $h$. This doesn’t negatively impact recall, as recall is measured by the percentage of actual contest joins that the model successfully recommends.

The online results show that WiDIR significantly outperforms other systems in terms of business metrics. By effectively learning player preferences, it drives an increase in contest joins, which subsequently improves CEAand GGR. While the LGB ranker also performs better than Popular in the CJ and CEA metrics, it performs poorly in GGR. 


\section{Conclusion \& Future scope}
\label{sec:conclusion}
We presented our model-wide and deep interaction ranker for recommending personalized contests at scale. Our online and offline experiments demonstrate that the model significantly improves key player relevance metrics, such as precision and recall, alongside critical business metrics like CJ, CEA, and GGR.

The proposed WiDIR (Wide and Deep Interaction Ranker) has demonstrated strong performance in personalized contest recommendations; however, several avenues remain for further enhancement. Future work will integrate edge models to optimize last-mile inference and incorporate session-based features, enabling real-time personalization closer to the player. The model’s interaction features provide inherent flexibility to adapt to evolving playing patterns. Additionally, we aim to explore multi-armed bandits to facilitate richer dynamic exploration-exploitation strategies, ensuring contest recommendations continuously evolve in response to changing player behaviour.

Beyond ranking optimization, multi-task learning presents an opportunity to extend WiDIR beyond a single-task ranker, making it adaptable to multiple personalization objectives. To further enhance representation learning, we plan to capture higher-order relationships in player-contest interactions by investigating graph modelling approaches, which could uncover deeper structural dependencies and improve recommendation diversity. Additionally, scalable online learning techniques will be explored to maintain efficiency while dynamically adjusting to the ever-changing contest landscape.


\begin{credits}

\subsubsection{\ackname} 
We would like to express our sincere gratitude to Nilesh Patil, Michael Zhao for their insightful feedback and guidance throughout this work. We also acknowledge the foundational contributions of Aditya Narisetty, Sriram Mudunuri, and others who laid the groundwork for this project. Special thanks to the Dream11 engineering teams for their assistance in setting up the inference layer, online feature store, and front-end integration, which were critical to the deployment and evaluation of our system.


\end{credits}

\begin{thebibliography}{1}
\providecommand{\url}[1]{\texttt{#1}}
\providecommand{\urlprefix}{URL }
\providecommand{\doi}[1]{https://doi.org/#1}

\bibitem{tensorflow2015-whitepaper}
Abadi, M., Agarwal, A., Barham, P., Brevdo, E., Chen, Z., Citro, C., Corrado, G.S., Davis, A., Dean, J., Devin, M., Ghemawat, S., Goodfellow, I., Harp, A., Irving, G., Isard, M., Jia, Y., Jozefowicz, R., Kaiser, L., Kudlur, M., Levenberg, J., Man\'{e}, D., Monga, R., Moore, S., Murray, D., Olah, C., Schuster, M., Shlens, J., Steiner, B., Sutskever, I., Talwar, K., Tucker, P., Vanhoucke, V., Vasudevan, V., Vi\'{e}gas, F., Vinyals, O., Warden, P., Wattenberg, M., Wicke, M., Yu, Y., Zheng, X.: {TensorFlow}: Large-scale machine learning on heterogeneous systems (2015), \url{https://www.tensorflow.org/}, software available from tensorflow.org

\bibitem{cheng2016wide}
Cheng, H.T., Koc, L., Harmsen, J., Shaked, T., Chandra, T., Aradhye, H., Anderson, G., Corrado, G., Chai, W., Ispir, M., et~al.: Wide \& deep learning for recommender systems. In: Proceedings of the 1st workshop on deep learning for recommender systems. pp. 7--10 (2016)

\bibitem{chugh2023ready}
Chugh, V., Kretz, W., Rosa, S., Beckelman, B., Tong, H., Mitkovski, D.: Ready for the big game: How fanduel boosts conversions and revenue at scale with a real-time recommender system for daily fantasy sports. In: 2023 IEEE International Conference on Big Data (BigData). pp. 1677--1686. IEEE (2023)

\bibitem{mordor_fantasy_sports}
Intelligence, M.: Fantasy sports market: Growth, trends, covid-19 impact, and forecasts (2023--2028). Online (2023), \url{https://www.mordorintelligence.com/industry-reports/fantasy-sports-market}, [Accessed 13 March 2025]

\bibitem{joachims2002optimizing}
Joachims, T.: {Optimizing Search Engines Using Clickthrough Data}. In: Proceedings of the Eighth ACM SIGKDD International Conference on Knowledge Discovery and Data Mining (KDD). pp. 133--142 (2002)

\bibitem{ke2017lightgbm}
Ke, G., Meng, Q., Finley, T., Wang, T., Chen, W., Ma, W., Ye, Q., Liu, T.Y.: Lightgbm: A highly efficient gradient boosting decision tree. Advances in neural information processing systems  \textbf{30} (2017)

\bibitem{zaharia2018accelerating}
Zaharia, M., Chen, A., Davidson, A., Ghodsi, A., Hong, S.A., Konwinski, A., Murching, S., Nykodym, T., Ogilvie, P., Parkhe, M., et~al.: Accelerating the machine learning lifecycle with mlflow. IEEE Data Eng. Bull.  \textbf{41}(4),  39--45 (2018)

\end{thebibliography}

\end{document}